\documentclass[french,english]{revtex4-1}
\usepackage[T1]{fontenc}
\usepackage[latin9]{inputenc}
\usepackage{textcomp}
\usepackage{amssymb}
\usepackage{esint}

\makeatletter
\usepackage[colorlinks = true,
            linkcolor = red,
            urlcolor  = blue,
            citecolor = blue,
            anchorcolor = blue]{hyperref}
\makeatother

\usepackage{babel}
\makeatletter
\addto\extrasfrench{%
   \providecommand{\fg}{\ifdim\lastskip>\z@\unskip\fi~\frqq}%
}

\makeatother
\begin{document}

\title[NC Continuity Eq with Nl Potential]{{\normalsize{}Continuity Equation in Presence of a Non-local potential
in Non-Commutative Phase-Space}}

\author{{\normalsize{}Ilyas Haouam}}

\email{ilyashaouam@live.fr, ilyashaouam@ymail.com}

\selectlanguage{english}%

\address{Laboratoire de Physique Mathématique et de Physique Subatomique (LPMPS),
Université Frères Mentouri, Constantine 25000, Algeria}
\begin{abstract}
{\normalsize{}We studied the continuity equation in presence of a
local potential, and a non-local potential arising from electron-electron
interaction in both commutative and non-commutative phase-space. Furthermore,
we examined the influence of the phase-space non-commutativity on
both the locality and the non-locality, where the definition of current
density in commutative phase-space cannot satisfy the condition of
current conservation, but with the steady state, in order to solve
this problem, we give a new definition of the current density including
the contribution due to the non-local potential. We showed that the
calculated current based on the new definition of current density
maintains the current. As well for the case when the non- commutativity
in phase-space considered, we found that the conservation of the current
density completely violated; and the non-commutativity is not suitable
for describing the current density in presence of non-local and local
potentials. Nevertheless, under some conditions, we modified the current
density to solve this problem. Subsequently, as an application we
studied the Frahn-Lemmer non-local potential, taking into account
that the employed methods concerning the phase-space non-commutativity
are both of Bopp-shift linear transformation through the Heisenberg-like
commutation relations, and the Moyal-Weyl product. }{\normalsize \par}

$\phantom{}$

$\phantom{}$

\textbf{Keywords:} Continuity equation; non-local potential; non-commutative
Schrödinger equation; phase-space non-commutativity; Frahn-Lemmer
potential; Moyal Product; Bopp-Shift Linear Transformation.

$\phantom{}$

\textbf{PACS numbers:} 03.65.Ge, 03.65.Pm ,0 02.40.Gh.
\end{abstract}

\keywords{Continuity equation; non-local potential; non-commutative Schrödinger
equation; phase-space non-commutativity; Frahn-Lemmer potential; Moyal
Product; Bopp-Shift Linear Transformation.}

\pacs{03.65.Ge, 03.65.Pm ,0 02.40.Gh.}

\maketitle

\section{{\normalsize{}Introduction}}

The power of physics resides in the fact of a single simple framework
can describe extremely different systems. But even the largest descriptive
equations sometimes reach their limits. A considered number of physics
equations are only approximations. What physicists really want is
not an approximation, but they want equations that connect the world's
behaviors directly to the foundations of reality, it's a big challenge.
But with the equations of motion, they can realize at least part of
that. While we always can extract the continuity equations from the
equations of motion, we can consider also that the continuity equation
is not an approximation but one of the equations that describe the
basic concepts, where it expresses the kinematical aspect of a symmetry
and is a useful auxiliary equation.

In the last years, in the development of nanotechnology, the transport
properties of nanodevices more and more become important, and it is
very interesting to understand how the current flows inside the nanodevices,
and how the current density gives the information about the heat dissipation
for example \cite{key-1}, as well as in the elementary particle physics,
for the quarks and the gluons which have a color charge \cite{key-2,key-3}
(in the theory of quantum chromodynamics (QCD)), in which it is always
conserved such as the electric charge. There is a continuity equation
for such color charge current (given at the gluon field strength tensor). 

There are a numerous quantities which are often or always conserved,
such as the baryon number which is proportional to the number of quarks
minus the number of antiquarks, also the lepton number, the isospin
(term used to describe groups of particles which have nearly the same
mass, such as the proton and the neutron)...etc, which means, in order
to investigate their own conservation laws we have to reach their
continuity equations. Where the continuity equation is an equation
that describes the transport of some quantity, furthermore there must
be a quantity $x$ which can flow or move (such as the energy, the
particle charge...), with $\rho_{x}$ being the volume density of
this quantity, the way this quantity flows is described by its flux,
denoted $\mathcal{J}_{x}$. Knowing that, the continuity equation
is another form of the conservation law, will be meaningful when it
is applied to a conserved quantity, and can be expressed in the integral
form, or in the differential form by the divergence theorem, it is
related always to the conservation of probability in quantum mechanics
(QM).

In this paper, we derive the continuity equation for a particle subjected
to a non-local and a local potential \cite{key-4} in non-commutative
phase-space (NCPS). Why do we care about the non-local potential and
the non-commutative geometry (NCG) ? The reason behind that, the Schrödinger
equation in the presence of a non-local potential has been the subject
of many investigations for several years \cite{key-5,key-6,key-7}
such as, in the calculations of the transport properties of the nano-devices
through the density functional theory and the Green\textquoteright s
function theory \cite{key-8,key-9} yonder, there are many cases where
the non-local potential is present. In the scattering theories of
nucleons and nuclei \cite{key-10}, the non-locality is generated
by the exchange interaction between the nucleons in the nucleus (considered
in Hartree-Fock type calculations) \cite{key-11}. In addition, the
nuclear optical potential describing the movement between colliding
nuclei is also non-local \cite{key-12,key-13}. In most nuclear structure
and reaction calculations yet, the non-locality has been assumed to
be small, and several approximation methods have been introduced in
the local potential model for reproducing its contributions. 

This work is realized in the NCG due to the importance and the advantages
of it in both of quantum mechanics and quantum fields, and generally
in physics today. Knowing that, the origin of the NCG is relative
to the search for topological spaces (\foreignlanguage{french}{C\textasteriskcentered -}algebras)
of functions which are replaced by non-commutative algebras, later
the NCG concept was rekindled by A. Connes and others \cite{key-14,key-15,key-16,key-17},
who theorized the idea of a differential structure in a non-commutative
framework, by studying and defining the cyclic cohomology. Where they
showed that the notion of differential calculations on varieties had
a non-commutative equivalent. Then, this type of geometry found a
great support by many mathematical results: Characterizations of commutative
von Neumann algebras, \foreignlanguage{french}{Gelfand-Naïmark} theorem
on \foreignlanguage{french}{C\textasteriskcentered -}algebras, cyclic
cohomology of \foreignlanguage{french}{$C^{\infty}(M)$} algebra,
K theory of \foreignlanguage{french}{C\textasteriskcentered -}algebras,
relations between Dirac operators and Riemannian metrics...

A non-commutative space theory replaces the non-commutativity of operators
associated to space-time coordinates with a deformation in the algebra
of functions defined on space-time, and the non-commutative version
of the field theory is obtained by changing the commutative theory
to the non-commutative one, this done by replacing ordinary fields
with non-commutative fields and ordinary products with Moyal -Weyl
products. To be more precise, N. Seiberg and E.Witten in the past
few years made their famous article \cite{key-18}, which is from
the most cited articles, encouraged a wide amount of interest in NCG,
which became the mainstream for a couple years.

It is worthwhile to mention that, the idea of NCPS is based essentially
on the Seiberg-Witten map, the Bopp's shift method and the Moyal-Weyl
product.

In this paper, our aim is not to solve equations but to focus on extracting
continuity equations. The plan of this paper is as follows : In section
3 we derive the continuity equation for non-local and local potentials,
taking as an application the Frahn-Lemmer non-local potential. In
section 4, we have the interesting result of our paper, which is represented
in the non-commutative continuity equation for non-local and local
potentials (with an application of the Frahn-Lemmer non-local potential).

\section{{\normalsize{}Review of the non-commutative geometry}}

The non-commutative geometry is the theory in which space may not
commute anymore. Let us consider the operators of coordinates and
momentum in a d dimensional non-commutative phase-space $x_{i}^{nc}$
and $p_{i}^{nc}$ respectively. Where the non-commutative phase-space
operators satisfy the Heisenberg-like commutation relations \cite{key-19}
\begin{equation}
\left[x_{i}^{nc},x_{j}^{nc}\right]=i\Theta_{ij},\,\left[p_{i}^{nc},p_{j}^{nc}\right]=i\eta_{ij},\,\left[x_{i}^{nc},p_{j}^{nc}\right]=i\hbar^{eff}\delta_{ij}\:(i,j=1,..,d),\label{eq:2}
\end{equation}
the effective Planck constant being
\begin{equation}
\hbar^{eff}=\hbar\left(1+\xi\right),\label{eq:3}
\end{equation}
where $\xi=\frac{Tr(\Theta\eta)}{4\hbar^{2}}$, the consistency condition
is $\xi\ll1$. With $\Theta_{ij}$, $\eta_{ij}$ are antisymmetric
real constant $(d\times d)$ matrices and $\delta_{ij}$ is the identity
matrix. Theoretical predictions for non-commutative systems (concerning
the non-commutative parameters) have been compared to experimental
data (the maximum absolute energy shifts allowed by the experiment),
leading to bounds on the noncommutative parameters \cite{key-19,key-20}
:

\begin{equation}
\Theta\approx4.10^{-40}m^{2},\quad\eta\approx1,76.10^{-61}Kg^{2}m^{2}s^{-2},\label{eq:3w}
\end{equation}
these above bounds will be suppressed due to the weak magnetic field
used in the experiments $B\approx5mG$.

$\vphantom{}$

Since the system in which we study the effects of non-commutativity,
is three dimensional, we limit our calculations to the following non-commutative
algebra 
\begin{equation}
\left[x_{i}^{nc},x_{j}^{nc}\right]=i\epsilon_{ijk}\Theta_{k},\,\left[p_{i}^{nc},p_{j}^{nc}\right]=i\epsilon_{ijk}\eta_{k},\,\left[x_{i}^{nc},p_{j}^{nc}\right]=i\hbar^{eff}\delta_{ij}\:(i,j,k=1,2,3),\label{eq:4w}
\end{equation}
\foreignlanguage{french}{ take into account that we neglect the uncertainty
relation between $x_{i}^{nc}$ and $p_{j}^{nc}$ . With $\Theta_{ij},\;\eta_{ij}$
are (3\texttimes 3) antisymmetric matrices, and $\epsilon_{ijk}$
is Levi-Civita symbol and the summation convention is used}. We have
$\epsilon_{123}=\epsilon_{231}=\epsilon_{312}=-\epsilon_{321}=-\epsilon_{132}=-\epsilon_{213}=1$,
if $i=j,j=k,\;\epsilon_{ijk}=0$, and $\Theta_{k},\:\eta_{k}$ are
the non-commutativity parameters. They are real-valued and antisymmetric
constant matrices with the dimension of $length{}^{2}$ and $momentum{}^{2}$,
respectively.

$\vphantom{}$

In the three dimensional commutative phase-space, the coordinates
$x_{i}$ and the momentum $p_{i}$ satisfy the usual canonical commutation
relations
\begin{equation}
\left[x_{i},x_{j}\right]=\left[p_{i},p_{j}\right]=0,\,\left[x_{i},p_{j}\right]=i\hbar\delta_{ij}\qquad(i,j=1,2,3).\label{eq:1}
\end{equation}

The non-commutative geometry Eq.(\ref{eq:2}) is described at the
level of fields and actions by the Moyal-Weyl product ($\star$-product)
\cite{key-21,key-22,key-23}. Let \foreignlanguage{french}{$f$ and
$g$} be two arbitrary functions from\foreignlanguage{french}{ $\mathcal{R}^{4}$.
We define }$\star$ product as follows
\begin{equation}
\begin{array}{c}
(f\star g)(x)=\exp[\frac{i}{2}\Theta_{ab}\partial_{x_{a}}\partial_{x_{b}}]f\left(x_{a}\right)g\left(x_{b}\right)=f(x)g(x)+\sum_{n=1}\left(\frac{1}{n!}\right)\left(\frac{i}{2}\right)^{n}\Theta^{a_{1}b_{1}}...\Theta^{a_{n}b_{n}}\partial_{a_{1}}...\partial_{a_{k}}f(x)\partial_{b_{1}}...\partial_{b_{k}}g(x).\end{array}\label{eq:4}
\end{equation}

Note that in our calculations, we use the following $\star$-product
properties : 

The Complex conjugation
\begin{equation}
\mbox{ }\left(f\left(x\right)\star g\left(x\right)\right)^{\ast}=g^{\ast}\left(x\right)\star f^{\ast}\left(x\right).\label{eq:6w}
\end{equation}

The $\star$-product under the integral sign
\begin{equation}
\mbox{ }\int\left(f\star g\right)\left(x\right)d^{4}x=\int\left(g\star f\right)\left(x\right)d^{4}x=\int\left(fg\right)\left(x\right)d^{4}x.\label{eq:7w}
\end{equation}

The non-commutative field theories for the low energies ($E<\frac{1}{\sqrt{\Theta}}$
) or the slowly varying fields effectively reduce to their commutative
version due to the nature of the $\star$-product.

The non-commutative phase-space operators are related to the commutative
phase-space one through the commutative Heisenberg-Weyl algebra in
terms of the known Bopp-shift\textbf{ }linear transformation which
was introduced from the Eq.(\ref{eq:4}) \cite{key-24,key-25}, and
it is given by

\begin{equation}
\begin{array}{cc}
x_{i}^{nc}=x_{i}-\frac{1}{2\hbar}\Theta_{ij}p_{j},\: & p_{i}^{nc}=p_{i}+\frac{1}{2\hbar}\eta_{ij}x_{j}\end{array}.\label{eq:5}
\end{equation}

If $\Theta=\eta=0$ the non-commutative phase-space framework will
become commutative one.

\selectlanguage{french}%

\section{{\normalsize{}Schrödinger equation in the presence of a nonlocal
potential in commutative phase-space }}

\selectlanguage{english}%
In presence of a non-local potential $V{}_{NL}(\mathbf{r},\mathbf{r}^{'})$,
the wave function obeys the following Schrödinger equation
\begin{equation}
\frac{\mathbf{p}^{2}}{2m}\psi(\mathbf{r},t)+\int V_{NL}(\mathbf{r},\mathbf{r}^{'})\psi(\mathbf{r}^{'},t)d\mathbf{r}^{'}=i\hbar\frac{\partial}{\partial t}\psi(\mathbf{r},t).\label{eq:7-1}
\end{equation}
\foreignlanguage{french}{A non-local potential operating on a wave
function \cite{key-26} has the form}
\begin{equation}
\int V_{NL}(\mathbf{r},\mathbf{r}^{'})\psi(\mathbf{r}^{'},t)d\mathbf{r}^{'}=\int V_{NL}(\mathbf{r},\mathbf{r}+\mathbf{s})\psi(\mathbf{r}+\mathbf{s},t)d\mathbf{s},\label{eq:8-1}
\end{equation}
with $\mathbf{r}^{'}=\mathbf{r}+\mathbf{s},$ and $d\mathbf{r}^{'}=d\mathbf{s}$,
taking into account that $Re[V_{NL}(\mathbf{r},\mathbf{r}^{'})]=Re[V_{NL}(\mathbf{r}^{'},\mathbf{r})]$
(symmetric), and using the Taylor series $\psi(\mathbf{r}+\mathbf{s})=(1+\mathbf{s}\frac{\partial}{\partial\mathbf{r}}+\frac{\mathbf{s}^{2}}{2}\frac{\partial^{2}}{\partial\mathbf{r}^{2}}+\frac{\mathbf{s}^{3}}{3!}\frac{\partial^{3}}{\partial\mathbf{r}^{\mathbf{3}}}+..)\psi(\mathbf{r})=e^{\frac{i\mathbf{sp}}{\hbar}}\psi(\mathbf{r})$,
we find 
\begin{equation}
\int V_{NL}(\mathbf{r},\mathbf{r}^{'})\psi(\mathbf{r}^{'},t)d\mathbf{r}^{'}=\int V_{NL}(\mathbf{r},\mathbf{r}+\mathbf{s})e^{\frac{i}{\hbar}\mathbf{sp}}d\mathbf{s}\psi(\mathbf{r},t),\label{eq:7}
\end{equation}
\foreignlanguage{french}{we can denote}
\begin{equation}
\int V_{NL}(\mathbf{r},\mathbf{r}+\mathbf{s})e^{\frac{i\mathbf{sp}}{\hbar}}d\mathbf{s}\psi(\mathbf{r},t)=\int dse^{\frac{i\mathbf{sp}}{2\hbar}}V_{NL}(\mathbf{r}-\frac{\mathbf{s}}{2},\mathbf{r}+\frac{\mathbf{s}}{2})e^{\frac{i\mathbf{sp}}{2\hbar}}\psi(\mathbf{r},t)=V_{NL}(\mathbf{r},\mathbf{p})\psi(\mathbf{r},t),\label{eq:8}
\end{equation}
then simply Eq.(\ref{eq:7-1}) takes the form
\begin{equation}
\frac{\mathbf{p}^{2}}{2m}\psi(\mathbf{r},t)+V_{NL}(\mathbf{r},\mathbf{p})\psi(\mathbf{r},t)=i\hbar\frac{\partial}{\partial t}\psi(\mathbf{r},t).\label{eq:10}
\end{equation}

\subsection{{\normalsize{}Frahn-Lemmer Non-local (Perey-Buck) Potential }}

In order to facilitate the calculations, and as an application concerning
the non-local potential, we choose the Frahn-Lemmer potential \cite{key-27}
defined as
\begin{equation}
\mathcal{V}_{NL}(\mathbf{r},\mathbf{r}^{'})=\mathcal{U}\left(\frac{1}{2}\left|\mathbf{r}+\mathbf{r}^{'}\right|\right)\mathcal{H}\left(\left|\mathbf{r}-\mathbf{r}^{'}\right|\right),\label{eq:11}
\end{equation}
where $\mathcal{U},\:\mathcal{H}$ represent the local average value
and the width of the non-locality respectively, knowing that, for
simplicity we consider $\mathcal{U}\left(\frac{1}{2}\left|\mathbf{r}+\mathbf{r}^{'}\right|\right)\approx V_{0}$,
and $\mathcal{H}$ should be a normalized Gaussian function as 
\begin{equation}
\mathcal{H}\left(\left|\mathbf{r}-\mathbf{r}^{'}\right|\right)=\frac{1}{\left(\pi\beta^{2}\right)^{\frac{3}{2}}}e^{-\frac{\left(\mathbf{r}-\mathbf{r}^{'}\right)^{2}}{\beta^{2}}},\label{eq:12}
\end{equation}
which is normalized so that
\begin{equation}
\int\mathcal{H}\left(\left|\mathbf{r}-\mathbf{r}^{'}\right|\right)d\mathbf{r}^{'}=1,\label{eq:16-1}
\end{equation}

\selectlanguage{french}%
then, our non-local potential takes the form {[}once the range parameter
is very small, the non-local potential tends to $\approx V_{0}\delta\left(\mathbf{r}-\mathbf{r}^{'}\right)${]}\foreignlanguage{english}{
\begin{equation}
\mathcal{V}_{NL}(\mathbf{r},\mathbf{r}^{'})=\frac{V_{0}}{\left(\pi\beta^{2}\right)^{\frac{3}{2}}}e^{-\frac{\left(\mathbf{r}-\mathbf{r}^{'}\right)^{2}}{\beta^{2}}},\label{eq:13}
\end{equation}
with $\beta,\;V_{0}$ are the non-locality range (typically takes
on a value of 0.85fm) and the depth of the nuclear potential (Wood-Saxon
function type) respectively.}

\selectlanguage{english}%
The following equation, obtained by putting Eq.(\ref{eq:13}) into
Eq.(\ref{eq:7-1})
\begin{equation}
\frac{\mathbf{p}^{2}}{2m}\psi(\mathbf{r},t)+\int\frac{V_{0}}{\left(\pi\beta^{2}\right)^{\frac{3}{2}}}e^{-\frac{\left(\mathbf{r}-\mathbf{r}^{'}\right)^{2}}{\beta^{2}}}\psi(\mathbf{r}^{'},t)d\mathbf{r}^{'}=i\hbar\frac{\partial}{\partial t}\psi(\mathbf{r},t),\label{eq:15}
\end{equation}
using Eq.(\ref{eq:8}), we have
\begin{equation}
\mathcal{V}_{NL}(\mathbf{r},\mathbf{p})=\int dse^{\frac{i\mathbf{sp}}{2\hbar}}\{\frac{V_{0}}{\left(\pi\beta^{2}\right)^{\frac{3}{2}}}e^{-\frac{\left(\mathbf{r}-\frac{\mathbf{s}}{2}-\mathbf{r}-\frac{\mathbf{s}}{2}\right)^{2}}{\beta^{2}}}\}e^{\frac{i\mathbf{sp}}{2\hbar}}=\int d\mathbf{s}\frac{V_{0}}{\left(\pi\beta^{2}\right)^{\frac{3}{2}}}e^{-\frac{\mathbf{s}^{2}}{\beta^{2}}+\frac{i}{\hbar}\mathbf{sp}},\label{eq:15-2}
\end{equation}
then, using the integral $\intop_{-\infty}^{+\infty}e^{-A^{2}\mathbf{s}^{2}+B\mathbf{s}}d\mathbf{s}=\frac{\sqrt{\pi}}{A}e^{\frac{\mathbf{B}^{2}}{4A^{2}}}$,
with $A=\frac{1}{\beta},\quad\mathbf{B}=\frac{i}{\hbar}\mathbf{p}$
, Eq.(\ref{eq:15}) turns to 
\begin{equation}
\frac{\mathbf{p}^{2}}{2m}\psi(\mathbf{r},t)+V_{0}e^{-\frac{\mathbf{p}^{2}\beta^{2}}{4\hbar^{2}}}\psi(\mathbf{r},t)=i\hbar\frac{\partial}{\partial t}\psi(\mathbf{r},t).\label{eq:20}
\end{equation}

What we need here is to show the form of the Schrödinger equation
in interaction with the Frahn-Lemmer non-local potential, knowing
that if we want to solve the above equation, we have to use the Fourier
transform to switch for the momentum representation (P representation),
there the calculations shall be very easy, knowing that the equation
will be time-independently considered, with $\mathbf{p}=\hbar\mathbf{k}$,
Eq.(\ref{eq:20}) becomes
\begin{equation}
E-\frac{\mathbf{\left(\hbar\mathbf{k}\right)}^{2}}{2m}=V_{0}e^{-\frac{\mathbf{k}^{2}\beta^{2}}{4}},\label{eq:23-1}
\end{equation}
this explains the connection between nuclear potential and momentum.
In another way, the strength of the potential decreases rapidly with
increasing momentum.

\subsection{{\normalsize{}The Continuity Equation in Commutative Phase-Space}}

The \foreignlanguage{french}{Schrödinger equation in presence of a
non-local potential $V_{NL}(\mathbf{r},\mathbf{r}^{'})$ and a local
potential $V_{L}(\mathbf{r})$ is written as follows
\begin{equation}
i\hbar\frac{\partial}{\partial t}\psi(\mathbf{r},t)=\frac{-\hbar^{2}}{2m}\mathbf{\nabla}^{2}\psi(\mathbf{r},t)+\int V_{NL}(\mathbf{r},\mathbf{r}^{'})\psi(\mathbf{r}^{'},t)d\mathbf{r}^{'}+V_{L}(\mathbf{r})\psi(\mathbf{r},t),\label{eq:25}
\end{equation}
the complex conjugate of the above equation written as
\begin{equation}
-i\hbar\frac{\partial}{\partial t}\psi^{\dagger}(\mathbf{r},t)=\frac{-\hbar^{2}}{2m}\mathbf{\nabla}^{2}\psi^{\dagger}(\mathbf{r},t)+\int V_{NL}^{\ast}(\mathbf{r},\mathbf{r}^{'})\psi^{\dagger}(\mathbf{r}^{'},t)d\mathbf{r}^{'}+V_{L}^{\ast}(\mathbf{r})\psi^{\dagger}(\mathbf{r},t),\label{eq:26}
\end{equation}
}here $\ast$ and $\dagger$ stand for the complex conjugation of
the potentials and for the wave-functions successively.

In order to find the continuity equation, \foreignlanguage{french}{we
have $\psi^{\dagger}(\mathbf{r},t)Eq$.(\ref{eq:25}) and $\psi(\mathbf{r},t)Eq$.(\ref{eq:26}),
so that we obtain
\begin{equation}
i\hbar\psi^{\dagger}(\mathbf{r},t)\frac{\partial}{\partial t}\psi(\mathbf{r},t)=\frac{-\hbar^{2}}{2m}\psi^{\dagger}(\mathbf{r},t)\mathbf{\nabla}^{2}\psi(\mathbf{r},t)+\int\psi^{\dagger}(\mathbf{r},t)V_{NL}(\mathbf{r},\mathbf{r}^{'})\psi(\mathbf{r}^{'},t)d\mathbf{r}^{'}+\psi^{\dagger}(\mathbf{r},t)V_{L}(\mathbf{r})\psi(\mathbf{r},t),\label{eq:27}
\end{equation}
\begin{equation}
-i\hbar\psi(\mathbf{r},t)\frac{\partial}{\partial t}\psi^{\dagger}(\mathbf{r},t)=\frac{-\hbar^{2}}{2m}\psi(\mathbf{r},t)\mathbf{\nabla}^{2}\psi^{\dagger}(\mathbf{r},t)+\int\psi(\mathbf{r},t)V_{NL}^{\ast}(\mathbf{r},\mathbf{r}^{'})\psi^{\dagger}(\mathbf{r}^{'},t)d\mathbf{r}^{'}+\psi(\mathbf{r},t)V_{L}^{\ast}(\mathbf{r})\psi^{\dagger}(\mathbf{r},t),\label{eq:28}
\end{equation}
}according to the subtraction of Eq.(\ref{eq:27}) from Eq.(\ref{eq:28})
we find
\begin{equation}
\begin{array}{c}
i\hbar\frac{\partial}{\partial t}\left(\psi^{\dagger}(\mathbf{r},t)\psi(\mathbf{r},t)\right)=\frac{-\hbar^{2}}{2m}\nabla\left(\psi^{\dagger}(\mathbf{r},t)\mathbf{\nabla}\psi(\mathbf{r},t)-\psi(\mathbf{r},t)\mathbf{\nabla}\psi^{\dagger}(\mathbf{r},t)\right)\\
+\int\left(\psi^{\dagger}(\mathbf{r},t)V_{NL}(\mathbf{r},\mathbf{r}^{'})\psi(\mathbf{r}^{'},t)-\psi(\mathbf{r},t)V_{NL}^{\ast}(\mathbf{r},\mathbf{r}^{'})\psi^{\dagger}(\mathbf{r}^{'},t)\right)d\mathbf{r}^{'}+\psi^{\dagger}(\mathbf{r},t)V_{L}(\mathbf{r})\psi(\mathbf{r},t)-\psi(\mathbf{r},t)V_{L}^{\ast}(\mathbf{r})\psi^{\dagger}(\mathbf{r},t)
\end{array},\label{eq:30}
\end{equation}
if we multiply the above equation by the charge carried by the particle,
we obtain the continuity equation of the corresponding particle, as
in the case of the electron, we multiply by $(-e)$. Eq.(\ref{eq:30})
may be contracted as
\begin{equation}
\frac{\partial\rho}{\partial t}+\mathbf{\nabla J}+\rho_{NL}+\rho_{L}=0,\label{eq:31}
\end{equation}
the obtained continuity equation Eq.(\ref{eq:31}) contains new quantities,
which are the non-local current density $\rho_{NL}$, and the local
density $\rho_{L}$, that is because of the consideration of the non-local
and the local interactions in the \foreignlanguage{french}{Schrödinger
} equation, where
\begin{equation}
\begin{array}{c}
\rho=J^{0}=\psi^{\dagger}(\mathbf{r},t)\psi(\mathbf{r},t)=\left|\psi(\mathbf{r},t)\right|^{2}\\
\mathbf{J}=\frac{-\hbar^{2}}{2m}\left(\psi^{\dagger}(\mathbf{r},t)\mathbf{\nabla}\psi(\mathbf{r},t)-\psi(\mathbf{r},t)\mathbf{\nabla}\psi^{\dagger}(\mathbf{r},t)\right)\\
\begin{array}{c}
\rho_{NL}=\int\left(\psi^{\dagger}(\mathbf{r},t)V_{NL}(\mathbf{r},\mathbf{r}^{'})\psi(\mathbf{r}^{'},t)-\psi(\mathbf{r},t)V_{NL}^{\ast}(\mathbf{r},\mathbf{r}^{'})\psi^{\dagger}(\mathbf{r}^{'},t)\right)d\mathbf{r}^{'}\\
\rho_{L}=\psi^{\dagger}(\mathbf{r},t)V_{L}(\mathbf{r})\psi(\mathbf{r},t)-\psi(\mathbf{r},t)V_{L}^{\ast}(\mathbf{r})\psi^{\dagger}(\mathbf{r},t)
\end{array}
\end{array}.\label{eq:32}
\end{equation}

If the local potential $V_{L}(\mathbf{r})$ is Hermitian, implying
that $\rho_{L}$ vanishes, which means that its symmetry maintained,
these results are similar to the calculations of Changsheng Li and
his colleagues \cite{key-28}.

In the steady state $\frac{\partial\rho}{\partial t}=0$, Eq.(\ref{eq:31})
becomes
\begin{equation}
\mathbf{\nabla J}+\rho_{NL}=0,\label{eq:32-1}
\end{equation}
if the non-local potential $V_{NL}(\mathbf{r},\mathbf{r}^{'})$ is
a real diagonal matrix, the quantity $\rho_{NL}$ vanishes, its symmetry
is maintained also. The current calculated from $\mathbf{J}$ is conserved
since $\mathbf{\nabla J}=0$. However, in presence of a non-local
potential, the quantity $\rho_{NL}$ is nonzero, and therefore $\mathbf{\nabla J}\neq0$.
As a result, the current calculated from the current density is not
conserved. Therefore, we need to modify the conventional definition
of the current density to include the contribution of $\rho_{NL}$
and $\rho_{L}$ induced by the non-local and the local potentials.

We define the new current density in the presence of a non-local and
a local potential as
\begin{equation}
\mathbf{J}_{tot}=\mathbf{J}+\mathbf{J}_{NL}+\mathbf{J}_{L},\label{eq:33}
\end{equation}
where $\mathbf{J}_{L}$, $\mathbf{J}_{NL}$ are the local current
density and the non-local current density (We call it the non-local
current density because it is merely due to the non-local potential)
defined as
\begin{equation}
\begin{array}{c}
\mathbf{J}_{NL}=-\nabla\chi_{NL}(\mathbf{r})\\
\mathbf{J}_{L}=-\nabla\varphi_{L}(\mathbf{r})
\end{array},\label{eq:34-1}
\end{equation}
where $\chi_{NL}(\mathbf{r})$, $\varphi_{L}(\mathbf{r})$ determined
by the following Poisson equation set
\begin{equation}
\begin{array}{c}
\nabla^{2}\chi_{NL}(\mathbf{r})+\rho_{NL}=0\\
\nabla^{2}\varphi_{L}(\mathbf{r})+\rho_{L}=0
\end{array},\label{eq:34-2}
\end{equation}
by solving each Poisson equation with proper boundary conditions,
we can calculate $\mathbf{J}_{NL}$ and $\mathbf{J}_{L}$. 

It is obvious that the newly defined current density satisfies $\nabla\mathbf{J}_{tot}=0$
and therefore, the calculated current from this current density satisfies
the current conservation.

Anywise, in the absence of the interactions, the continuity equation
takes its simple known form in the quantum mechanics
\begin{equation}
i\hbar\frac{\partial}{\partial t}\left|\psi(\mathbf{r},t)\right|^{2}+\frac{\hbar^{2}}{2m}\nabla\left(\psi^{\dagger}(\mathbf{r},t)\mathbf{\nabla}\psi(\mathbf{r},t)-\psi(\mathbf{r},t)\mathbf{\nabla}\psi^{\dagger}(\mathbf{r},t)\right)=0.\label{eq:34}
\end{equation}

\selectlanguage{french}%

\section{{\normalsize{}Schrödinger equation in presence of a nonlocal potential
in non-commutative phase-space }}

We introduce the non-commutativity in space through the $\star$-product,
the Schrödinger equation in presence of non-local and local potentials
in non-commutative space is written as

\selectlanguage{english}%
\begin{equation}
\frac{\mathbf{p}^{2}}{2m}\psi(\mathbf{r},t)+\int V_{NL}(\mathbf{r},\mathbf{r}^{'})\star\psi(\mathbf{r}^{'},t)d\mathbf{r}^{'}+V{}_{L}(\mathbf{r})\star\psi(\mathbf{r},t)=i\hbar\frac{\partial}{\partial t}\psi(\mathbf{r},t),\label{eq:35}
\end{equation}
taking into account that t\foreignlanguage{french}{he $\star$-product}
under the integral sign become ordinary product as shown in Eq.(\ref{eq:7w}),
leading to found out that the symmetry of the real part of the non-local
potential maintained in the non-commutative framework.

For $V_{L}(\mathbf{r})\sim hr$, with $h$ is real-valued,  and using
Eq.(\ref{eq:4}) we find
\begin{equation}
V{}_{L}(\mathbf{r})\star\psi(\mathbf{r},t)=V{}_{L}(\mathbf{r})\psi(\mathbf{r},t)+i\Theta_{ab}\partial_{a}V{}_{L}(\mathbf{r})\partial_{b}\psi(\mathbf{r},t)+0\left(\Theta^{2}\right).\label{eq:37}
\end{equation}

\selectlanguage{french}%
Then we introduce the non-commutativity in phase by the mapping \foreignlanguage{english}{$\mathbf{p}\longrightarrow\mathbf{p^{nc}}$
through Bopp-shift translation Eq.(\ref{eq:5}), we have}

\selectlanguage{english}%
\begin{equation}
\frac{(\mathbf{p}^{nc})^{2}}{2m}\psi(\mathbf{r},t)+\int V_{NL}(\mathbf{r},\mathbf{r}^{'})\star\psi(\mathbf{r}^{'},t)d\mathbf{r}^{'}+V{}_{L}(\mathbf{r})\star\psi(\mathbf{r},t)=i\hbar\frac{\partial}{\partial t}\psi(\mathbf{r},t),\label{eq:11-1}
\end{equation}
where
\begin{equation}
(\mathbf{p}^{nc})^{2}=(p_{i}+\frac{1}{2\hbar}\eta_{ij}r_{j})^{2}=p_{i}^{2}+\frac{1}{2\hbar}\eta_{ij}p_{i}r_{j}+\frac{1}{2\hbar}\eta_{ij}r_{j}p_{i}+\frac{1}{4\hbar^{2}}\eta_{ij}\eta_{ik}r_{j}r_{k},\label{eq:42-1-1}
\end{equation}
we restrict ourselves only to the 1st order of the non-commutativity
in phase $0\left(\eta^{2}\right)$, {[}for the equilibrium with the
non-commutativity in space considered in this work{]}.

With $\eta_{ij}=-\eta_{ji}=\eta\epsilon_{ij}$, and $\eta_{k}=\frac{1}{2}\epsilon_{kij}\eta_{ij}\longrightarrow\eta_{ij}=2\eta_{k}\epsilon_{kij}$,
knowing that $(\epsilon_{kij})^{2}=1$, and $\left(U\times V\right)_{k}=\epsilon_{kij}U_{i}V_{j}$,
$\mathbf{L}=\mathbf{r}\times\mathbf{p}$ then

\begin{equation}
\begin{array}{cccccccc}
\frac{1}{2\hbar}\eta_{ij}p_{i}r_{j} & = & \frac{1}{\hbar}\eta_{k}\epsilon_{kij}p_{i}r_{j} & = & \frac{1}{\hbar}\left(\mathbf{p}\times\mathbf{r}\right)\mathbf{\eta} & = & -\frac{1}{\hbar}\mathbf{L}\mathbf{\eta}\\
\frac{1}{2\hbar}\eta_{ij}r_{j}p_{i} & = & -\frac{1}{2\hbar}\eta_{ji}r_{j}p_{i} & = & -\frac{1}{\hbar}\eta_{k}\epsilon_{kji}r_{j}p_{i} & = & -\frac{1}{\hbar}\left(\mathbf{r}\times\mathbf{p}\right)\mathbf{\eta} & =-\frac{1}{\hbar}\mathbf{L}\mathbf{\eta}
\end{array},\label{eq:35-1}
\end{equation}
substituting the above relations in Eq. (\ref{eq:42-1-1}). Finally,
we obtain

\begin{equation}
(\mathbf{p}^{nc})^{2}=\mathbf{p}^{2}-\frac{2}{\hbar}\mathbf{L}\mathbf{\eta}+0\left(\eta^{2}\right).\label{eq:38}
\end{equation}

Substituting Eqs.(\ref{eq:37}-\ref{eq:38}) in Eq.(\ref{eq:11-1}),
we obtain
\begin{equation}
i\hbar\frac{\partial}{\partial t}\psi(\mathbf{r},t)=\frac{-\hbar^{2}}{2m}\mathbf{\nabla}^{2}\psi(\mathbf{r},t)-\frac{1}{m\hbar}\mathbf{L}\mathbf{\eta}\psi(\mathbf{r},t)+\int V_{NL}(\mathbf{r},\mathbf{r}^{'})\psi(\mathbf{r}^{'},t)d\mathbf{r}^{'}+V{}_{L}(\mathbf{r})\psi(\mathbf{r},t)+i\Theta_{ab}\partial_{a}V{}_{L}(\mathbf{r})\partial_{b}\psi(\mathbf{r},t).\label{eq:39}
\end{equation}

\subsection{{\normalsize{}The Frahn-Lemmer Non-local Potential in Non-Commutative
Phase-Space}}

Knowing that$\int V_{NL}(\mathbf{r},\mathbf{r}^{'})\psi(\mathbf{r}^{'},t)d\mathbf{r}^{'}$
goes to $V_{NL}(\mathbf{r},\mathbf{p})\psi(\mathbf{r},t)$, the\foreignlanguage{french}{
Schrödinger equation in interaction with Frahn-Lemmer non-local potential
and a local potential in non-commutative phase-space is given by,}
\begin{equation}
\frac{-\hbar^{2}}{2m}\mathbf{\nabla}^{2}\psi(\mathbf{r},t)-\frac{1}{m\hbar}\mathbf{L}\mathbf{\eta}\psi(\mathbf{r},t)+V_{0}e^{-\frac{\left(\mathbf{p}^{2}-\frac{2}{\hbar}\mathbf{L}.\mathbf{\eta}\right)\beta^{2}}{4\hbar^{2}}}\psi(\mathbf{r},t)+V{}_{L}(\mathbf{r})\psi(\mathbf{r},t)+i\Theta_{ab}\partial_{a}V{}_{L}(\mathbf{r})\partial_{b}\psi(\mathbf{r},t)=i\hbar\frac{\partial}{\partial t}\psi(\mathbf{r},t),\label{eq:41}
\end{equation}
for $\left(\mathbf{p}^{2}-\frac{2}{\hbar}\mathbf{L}\mathbf{\eta}\right)\ll\frac{4\hbar^{2}}{\beta^{2}}$,
let us approximate as
\begin{equation}
V_{0}e^{-\frac{\left(\mathbf{p}^{2}-\frac{2}{\hbar}\mathbf{L}.\mathbf{\eta}\right)\beta^{2}}{4\hbar^{2}}}=V_{0}[1-\frac{\mathbf{p}^{2}\beta^{2}}{4\hbar^{2}}+\frac{\mathbf{L}\mathbf{\eta}\beta^{2}}{2\hbar^{3}}],\label{eq:42}
\end{equation}
with $a=\frac{\hbar^{2}}{2m}+\frac{V_{0}\beta^{2}}{4}$, $b=\frac{V_{0}\beta^{2}}{2\hbar^{3}}-\frac{1}{m\hbar}$
and substituting Eq.(\ref{eq:42}) into Eq.(\ref{eq:41}), we obtain
\begin{equation}
-a\mathbf{\nabla}^{2}\psi(\mathbf{r},t)+b\mathbf{L}\mathbf{\eta}\psi(\mathbf{r},t)+V_{0}\psi(\mathbf{r},t)+V{}_{L}(\mathbf{r})\psi(\mathbf{r},t)+i\Theta_{ab}\partial_{a}V{}_{L}(\mathbf{r})\partial_{b}\psi(\mathbf{r},t)=i\hbar\frac{\partial}{\partial t}\psi(\mathbf{r},t).\label{eq:43}
\end{equation}

The above equation is the non-commutative Schrödinger \foreignlanguage{french}{equation
in interaction with Frahn-Lemmer non-local potential and a local potential.
The non-commutativity in space influenced the local part, while the
noncommutativity in phase touched the non-local part.}

\subsection{{\normalsize{}The Continuity Equation in Non-Commutative Phase-Space}}

\selectlanguage{french}%
The Schrödinger equation in presence of a non-local and a local potential
in non-commutative phase-space is given by the Eq.(\ref{eq:39}),
and its complex conjugate is given by
\begin{equation}
\begin{array}{c}
-i\hbar\frac{\partial}{\partial t}\psi^{\dagger}(\mathbf{r},t)=\frac{-\hbar^{2}}{2m}\mathbf{\nabla}^{2}\psi^{\dagger}(\mathbf{r},t)-\frac{1}{m\hbar}\mathbf{L}\mathbf{\eta}\psi^{\dagger}(\mathbf{r},t)+\int V_{NL}^{\ast}(\mathbf{r},\mathbf{r}^{'})\psi^{\dagger}(\mathbf{r}^{'},t)d\mathbf{r}^{'}+V_{L}^{\ast}(\mathbf{r})\psi^{\dagger}(\mathbf{r},t)-i\Theta_{ab}\partial_{a}V_{L}^{\ast}(\mathbf{r})\partial_{b}\psi^{\dagger}(\mathbf{r},t),\end{array}\label{eq:44}
\end{equation}
from the multiplications $\psi^{+}(\mathbf{r},t)Eq$.(\ref{eq:39})
and $\psi(\mathbf{r},t)Eq$.(\ref{eq:44}), it comes
\begin{equation}
\begin{array}{c}
i\hbar\psi^{\dagger}(\mathbf{r},t)\frac{\partial}{\partial t}\psi(\mathbf{r},t)=\frac{-\hbar^{2}}{2m}\psi^{\dagger}(\mathbf{r},t)\mathbf{\nabla}^{2}\psi(\mathbf{r},t)-\frac{1}{m\hbar}\psi^{\dagger}(\mathbf{r},t)\mathbf{L}\mathbf{\eta}\psi(\mathbf{r},t)+\int\psi^{\dagger}(\mathbf{r},t)V_{NL}(\mathbf{r},\mathbf{r}^{'})\psi(\mathbf{r}^{'},t)d\mathbf{r}^{'}\\
+\psi^{\dagger}(\mathbf{r},t)V_{l}(\mathbf{r}).\psi(\mathbf{r},t)+i\Theta_{ab}\psi^{\dagger}(\mathbf{r},t)\partial_{a}V_{L}(\mathbf{r})\partial_{b}\psi(\mathbf{r},t)
\end{array},\label{eq:45}
\end{equation}

\begin{equation}
\begin{array}{c}
-i\hbar\psi(\mathbf{r},t)\frac{\partial}{\partial t}\psi^{\dagger}(\mathbf{r},t)=\frac{-\hbar^{2}}{2m}\psi(\mathbf{r},t)\mathbf{\nabla}^{2}\psi^{\dagger}(\mathbf{r},t)-\frac{1}{m\hbar}\psi(\mathbf{r},t)\mathbf{L}\mathbf{\eta}\psi^{\dagger}(\mathbf{r},t)+\int\psi(\mathbf{r},t)V_{NL}^{\ast}(\mathbf{r},\mathbf{r}^{'})\psi^{\dagger}(\mathbf{r}^{'},t)d\mathbf{r}^{'}\\
+\psi(\mathbf{r},t)V_{L}^{\ast}(\mathbf{r})\psi^{\dagger}(\mathbf{r},t)-i\Theta_{ab}\psi(\mathbf{r},t)\partial_{a}V_{L}^{\ast}(\mathbf{r})\partial_{b}\psi^{\dagger}(\mathbf{r},t),
\end{array},\label{eq:46}
\end{equation}
then, by the subtraction of Eq.(\ref{eq:45}) from Eq.(\ref{eq:46})
we obtain\foreignlanguage{english}{
\begin{equation}
\begin{array}{c}
i\hbar\frac{\partial}{\partial t}\left(\psi^{\dagger}(\mathbf{r},t)\psi(\mathbf{r},t)\right)=\frac{-\hbar^{2}}{2m}\nabla\left(\psi^{\dagger}(\mathbf{r},t)\mathbf{\nabla}\psi(\mathbf{r},t)-\psi(\mathbf{r},t)\mathbf{\nabla}\psi^{\dagger}(\mathbf{r},t)\right)+\frac{1}{m\hbar}\left(\psi(\mathbf{r},t)\mathbf{L}\mathbf{\eta}\psi^{\dagger}(\mathbf{r},t)-\psi^{\dagger}(\mathbf{r},t)\mathbf{L}\mathbf{\eta}\psi(\mathbf{r},t)\right)\\
+\int\left(\psi^{\dagger}(\mathbf{r},t)V_{NL}(\mathbf{r},\mathbf{r}^{'})\psi(\mathbf{r}^{'},t)-\psi(\mathbf{r},t)V_{NL}^{\ast}(\mathbf{r},\mathbf{r}^{'})\psi^{\dagger}(\mathbf{r}^{'},t)\right)d\mathbf{r}^{'}+\psi^{\dagger}(\mathbf{r},t)V{}_{L}(\mathbf{r})\psi(\mathbf{r},t)-\psi(\mathbf{r},t)V_{L}^{\ast}(\mathbf{r})\psi^{\dagger}(\mathbf{r},t)\\
+i\Theta_{ab}\left(\psi^{\dagger}(\mathbf{r},t)\partial_{a}V{}_{L}(\mathbf{r})\partial_{b}\psi(\mathbf{r},t)+\psi(\mathbf{r},t)\partial_{a}V_{L}^{\ast}(\mathbf{r})\partial_{b}\psi^{\dagger}(\mathbf{r},t)\right)
\end{array},\label{eq:47}
\end{equation}
contracting the above equation as follows
\begin{equation}
\frac{\partial\rho}{\partial t}+\mathbf{\nabla J}+\rho_{NL}+\rho_{L}^{nc}+\mathcal{C}^{nc}=0.\label{eq:48}
\end{equation}
}

\selectlanguage{english}%
Eq.(\ref{eq:47}) will be recognized as the non-commutative continuity
equation, denoting the separate terms in it as follows
\begin{equation}
\begin{array}{c}
\rho=\mathcal{J}{}^{0}=\psi^{\dagger}(\mathbf{r},t)\psi(\mathbf{r},t)=\left|\psi(\mathbf{r},t)\right|^{2}\\
\mathcal{\mathbf{J}}=\frac{-\hbar^{2}}{2m}\left(\psi^{\dagger}(\mathbf{r},t)\mathbf{\nabla}\psi(\mathbf{r},t)-\psi(\mathbf{r},t)\mathbf{\nabla}\psi^{\dagger}(\mathbf{r},t)\right)\\
\rho_{NL}=\int\left(\psi^{\dagger}(\mathbf{r},t)V_{NL}(\mathbf{r},\mathbf{r}^{'})\psi(\mathbf{r}^{'},t)-\psi(\mathbf{r},t)V_{NL}^{\ast}(\mathbf{r},\mathbf{r}^{'})\psi^{\dagger}(\mathbf{r}^{'},t)\right)d\mathbf{r}^{'}\\
\rho_{L}^{nc}=\psi^{\dagger}(\mathbf{r},t)V_{L}(\mathbf{r})\psi(\mathbf{r},t)-\psi(\mathbf{r},t)V_{L}^{\ast}(\mathbf{r})\psi^{\dagger}(\mathbf{r},t)+i\Theta_{ab}\left(\psi^{\dagger}(\mathbf{r},t)\partial_{a}V_{L}(\mathbf{r})\partial_{b}\psi(\mathbf{r},t)-\psi(\mathbf{r},t)\partial_{a}V_{L}^{\ast}(\mathbf{r})\partial_{b}\psi^{\dagger}(\mathbf{r},t)\right)\\
\mathcal{C}^{nc}=\frac{1}{m\hbar}\left(\psi(\mathbf{r},t)\mathbf{L}\mathbf{\eta}\psi^{\dagger}(\mathbf{r},t)+\psi^{\dagger}(\mathbf{r},t)\mathbf{L}\mathbf{\eta}\psi(\mathbf{r},t)\right)
\end{array}\mathbf{.}\label{eq:49}
\end{equation}

It is obvious that the conservation of the current density in the
non-commutative phase-space completely violated, which means that
the current density does not satisfy the current conservation. Then
we move to the interpretation of the separating terms, the existence
of the quantities corresponding to the explicit $\Theta,\:\eta$ parameters,
which are involved in the obtained equation Eq.(\ref{eq:49}) due
to the effect of the phase-space non-commutativity on \foreignlanguage{french}{the
Schrödinger equation.} Firstly, these quantities emerged merely as
terms containing the parameters $\Theta,\:\eta$ , consequently after
extracting the non-commutative continuity equation, those terms being
responsible for generating the new quantities collectively with the
correction term which contains $\Theta$ parameter.

More accurately, the effect of the non-local potential on the continuity
equation arises as a non-local quantity of density type, as well as
for the locality effect, it appears as a local quantity of density
type also, where the non-commutativity in phase formed only a correction
term $\mathcal{C}^{nc}$, which appeared in the non-commutative continuity
equation, but for the non-commutativity in space affected only the
local quantity $\rho_{L}^{nc}$ through a first-order correction.
Once the local potential is null, the local density quantity with
its non-commutative correction will disappear.

Comparing the continuity equation in commutative and in non-commutative
cases, we find that the non-commutativity influence is very clear
in the amount of the local potential, but for the non-locality amount
vanishes, we find that the non-commutativity effect violates the conservation
of the continuity equation.

In what follows, we modify the expression of the density current,
in which, it'll be conserved in the non-commutative phase-space:

If the local potential $V_{L}(\mathbf{r})$ is real, the quantity
$\rho_{L}^{nc}$ vanishes (similar to the commutative case), also
when the non-local potential $V_{NL}(\mathbf{r},\mathbf{r}^{'})$
is a real diagonal matrix, the quantity $\rho_{NL}$ vanishes. But
the current calculated from $\mathbf{J}$ is not conserved due to
the phase non-commutativity correction $\mathcal{C}^{nc}$. However,
while $\rho_{NL}$ and $\rho_{L}$ are nonzero. As a result, the current
density not conservable, as well as of the symmetry isn't maintained
in the non-commutative phase-space. Therefore, we need to modify the
conventional definition of the current density to include the contribution
of $\rho_{NL}$ and $\rho_{L}^{nc}$ , and in the steady state $\frac{\partial\rho}{\partial t}=0$,
Eq.(\ref{eq:48}) becomes
\begin{equation}
\mathbf{\nabla J}+\mathcal{C}^{nc}+\rho_{L}^{nc}+\rho_{NL}=0,\label{eq:32-1-1}
\end{equation}
we make the following replacement 
\begin{equation}
\mathbf{J}+\mathcal{\mathbf{\kappa}}^{nc}\rightarrow\mathcal{J}^{nc},\mbox{ with }\mathcal{C}^{nc}=\nabla\mathbf{\mathcal{\kappa}}^{nc},\label{eq:50}
\end{equation}
with the condition $\mathbf{\nabla}\mathcal{C}^{nc}=0$. We define
the new global current density in the presence of a non-local and
a local potential where the non-commutativity is considered as
\begin{equation}
\mathbf{J}_{tot}^{nc}=\mathcal{J}^{nc}+\mathbf{J}_{NL}+\mathbf{J}_{L}^{nc},\label{eq:33-1}
\end{equation}
where $\mathbf{J}_{L}^{nc}$, $\mathbf{J}_{NL}$ are the non-commutative
local current density, and the non-local current density defined as
\begin{equation}
\begin{array}{c}
\mathbf{J}_{NL}=-\nabla\chi_{NL}(\mathbf{r}),\;\mathbf{J}_{L}^{nc}=-\nabla\varphi_{L}^{nc}(\mathbf{r})\end{array},\label{eq:34-1-1}
\end{equation}
where $\chi_{NL}(\mathbf{r})$, $\varphi_{L}^{nc}(\mathbf{r})$ determined
by the following Poisson equation set
\begin{equation}
\begin{array}{c}
\nabla^{2}\chi_{NL}(\mathbf{r})+\rho_{NL}=0\\
\nabla^{2}\varphi_{L}^{nc}(\mathbf{r})+\rho_{L}^{nc}=0
\end{array},\label{eq:34-2-1}
\end{equation}
by solving each Poisson equation through the proper boundary conditions,
we obtain $\mathbf{J}_{NL}$ and $\mathbf{J}_{L}^{nc}$. Therefore
the newly defined total current density satisfies $\nabla\mathbf{J}_{tot}^{nc}=0$
and therefore, the calculated non-commutative current from this current
density satisfies the current conservation.

\section{{\normalsize{}Conclusion}}

In conclusion, the phase-space non-commutativity introduced in the
Schrödinger equation and consequently, the continuity equation obtained
in the case of commutativity and in the case of non-commutativity,
without forgetting that the Schrödinger equation considered in interaction
with non-local and local potentials, this, in turn, being responsible
for causing new quantities of density type in the continuity equation.
We found that the non-commutativity in phase-space is not suitable
for describing the current density in the presence of non-local and
local potentials.

Knowing that the phase-space non-commutativity effect introduced through
both of the Bopp-shift linear translation method and the Moyal-Weyl
product. Under the condition that space-space and momentum-momentum
are all commutative, the results in non-commutative phase-space return
to that of the usual quantum mechanics.

The results of the present work can be used to investigate the conservation
laws by involving the non-commutative geometry such as : the non-commutative
CPT symmetry (with the Lorentz invariance), the conservation of weak
isospin (with SU(2), Gauge invariance). In the electromagnetism also
such as the non-commutative Maxwell's equations, maybe also used in
the non-commutative general relativity. We are out looking to investigate
the Klein-paradox depending on these results.

\end{document}